\documentclass[12pt]{article}  

\usepackage{amsmath,amsfonts,amssymb}
\usepackage{graphicx}
\usepackage{float}
\usepackage[colorlinks=true, allcolors=blue]{hyperref}
\usepackage{makecell}

\providecommand{\keywords}[1]
{
  \small	
  \textbf{\textit{Keywords---}} #1
}

\title{Experimental study of the first telescope with a toroidal curved detector}
\author{Eduard Muslimov$^{a,b}$,Simona Lombardo$^{b}$,\\
Thibault Behaghel$^{b}$, Jiawei Liu$^{b}$, and Emmanuel Hugot$^{b,c}$\\
\small a--Department of Physics, University of Oxford, \\
\small Keble Rd, OX14 3RH Oxford, UK\\
\small b--Aix Marseille Univ, CNRS, CNES, LAM, Marseille, France\\
\small c -- CURVE s.a.s./Curve-One, Marseille, France\\
}

\begin{document} 
\maketitle{}

\begin{abstract}
We describe a practical implementation of the anamorphically curved detector concept. In order to demonstrate its advantages {,} a telescope lab prototype was developed, built {,} and tested. It is based on a 4-mirror all-spherical unobscured design, similar to that proposed by D. Shafer. The telescope is open at $F/\#=5.5$ and its extended field of view is $10.6^\circ \times 8^\circ$. We explore the design parameter space to demonstrate the change in gain introduced by the curved detector and  {to} substantiate the chosen parameters. If the image surface is curved to the shape of a toroid, the image quality reaches the diffraction limit over the entire field by design. The design was optimized to use standard concave  {spherical mirrors}. The detector was curved down toroidally with an exquisite surface precision of $\approx 11$  {micrometers} Peak-to-Valley. The experimental tests of the prototype  {have shown} that use of the toroidally curved detector allows to increase the contrast in the field corner by $\approx 0.3$ units, which is in good agreement with the  {modeling} results. We also propose a few prospective applications of the demonstrated concept.
\end{abstract}


\section{Introduction}
\label{sect:intro}  
The technology of curved detectors was proposed for the first time more than two decades ago  \cite{Nikzad01} as a direct solution  {to} the field curvature problem, which is a  {well-known} factor limiting optical system performance.  {Since then,} the concept has been actively developed and  {has had} a number of practical implementations {, for example,} in the works of Nikzad, Ko, Itonaga, Guenter, Dumas to cite a few  \cite{Ko08, Rim08, Dinyari08, Nikzad10, Itonaga11, Dumas12, Tekaya14, Gregory15, Hatakeyama16, Guenter17}. In some of these examples {,} the curved detectors were not only produced at the level of functional prototype but also applied in a real instrument. However, we believe that the realm of  {curved focal plane} optical instruments remains under-explored to a large extent.

In particular, previously the authors have shown  \cite{Muslimov18} that off-axis optical systems have an intrinsically anamorphic field curvature. The exact shape of the focal surface in  {an} off-axis reflective system was further investigated by Liu \cite{Liu24}.  {In both cases}, it was shown with computation and  {modeling} that the departure of the actual focal surface from the Petzval sphere may have a significant impact on the image quality over an extended field of view.  {In addition}, it was demonstrated by finite element analysis  that a complex shape of the detector surface could be generated with the existing curving process or a slightly modified version  {of it} without producing an extreme mechanical stress. Therefore, from the standpoint of design and  {modeling} an uncountable number of options should become possible with the anamorphic and even general freeform curved sensors. The next step would be demonstrating that the precise machining and bending technologies as well as  {measurement} and alignment accuracy would not prevent us from implementing this concept in practice.  So, the main goal of the present study is to provide an experimental proof  {of} the proposed design approach and  {the} corresponding technical solutions.  

To reach this goal, we were to solve the following tasks, driving the paper structure:
\begin{itemize}
    \item Choose the optical design type, which could demonstrate the effect of the anamorphically curved detector surface. Find such a combination of its main parameters that would make the difference in image quality measurable, but allow us to keep the design relatively simple and feasible (Sec.~\ref{sec:design}).
    \item Manufacture the toroidal curved detectors and control their surface shape and functionality (Sec.~\ref{sec:detector}).
    \item Develop an experiment demonstrating the gain in imaging performance (Sec.~\ref{sec:experiments}).
    \item Summarize the results and propose prospective applications and further developments (Sec.~\ref{sec:conc}). 
    
\end{itemize}

\section{Optical design and analysis}
\label{sec:design}
First, we should choose the basic optical design for the future demonstrator. It has to satisfy a few key criteria, namely:
\begin{itemize}
    \item The optical system should have an off-axis geometry generating an anamorphic field curvature. There is a wide variety of unobscured catoptric and catadioptric designs  {that} match this requirement.
    \item It should have the image quality allowing to visually demonstrate the gain obtained due to the use of an anamorphically curved focal  {surface}.  {This} implies that  {all aberrations} except for the field curvature should be well corrected. It is useful to note here that we are not limited by the nominal field of view and focal ratio ($F/\#$) of any prototype design.
    \item The optical design should be relatively simple.  {This} means that a priority should be given to options with  {fewer} optical components and simpler surfaces. Also, it is useful to set limits on the overall dimensions  {of the system}. 
\end{itemize}

\begin{figure}[htbp]
\centering
\fbox{\includegraphics[width=0.8\linewidth]{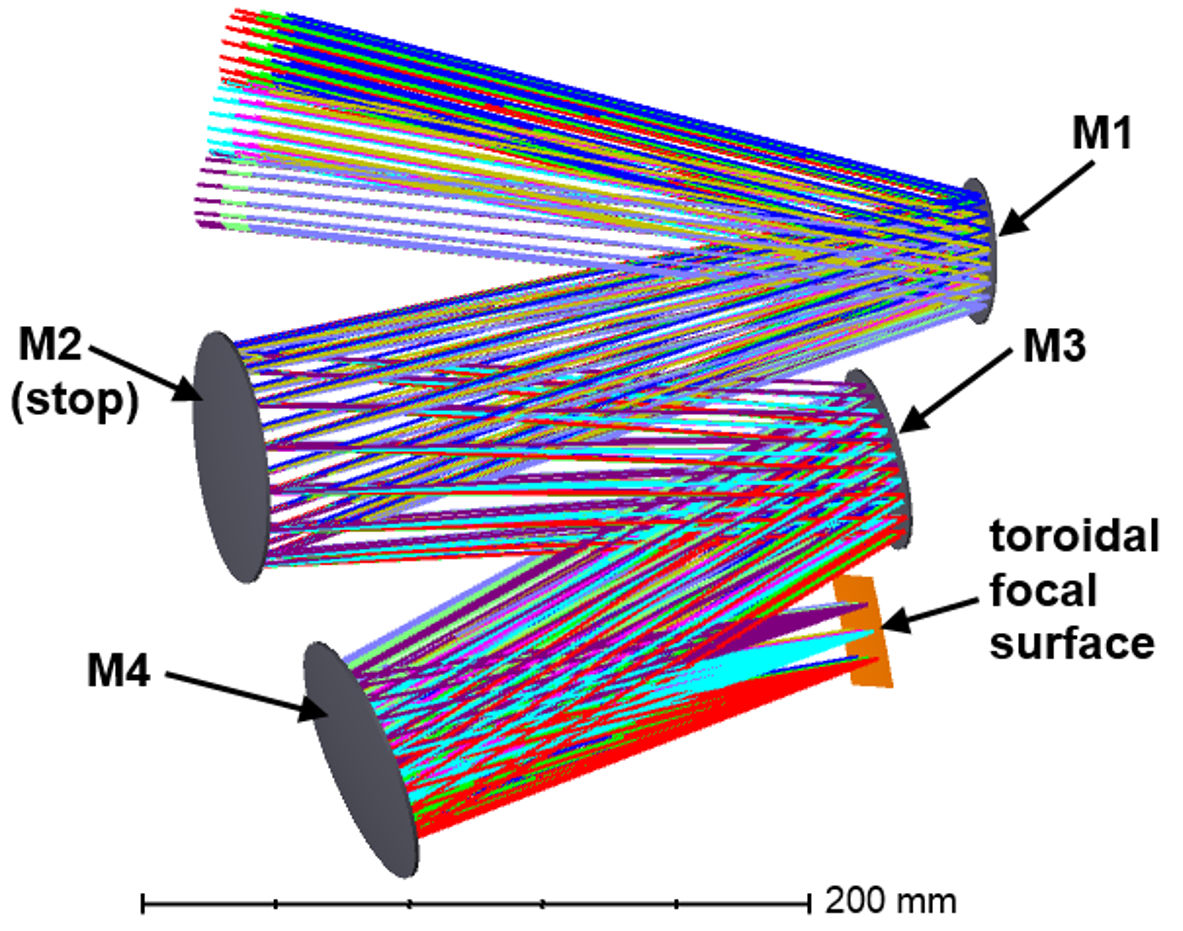}}
\caption{The all-reflective unobscured telescope optical design.}
\label{fig:Shafer_design}
\end{figure}

A number of elegant all-spherical unobscured telescope designs were proposed by David Shafer in 1978 and later became a textbook example for unobscured telescope designs \cite{Shafer78, Bass94} - see Fig.~\ref{fig:Shafer_design} for its general view. This four-mirror telescope design satisfies all the conditions given above,  {since} it uses only spherical surfaces (2 convex and 2 concave ones) and provides a  {diffraction-limited} image quality for $F/\#=4.5$ at {$1064 nm$}.  {The field curvature in the original design is compensated to some extent due to combination of concave and convex surfaces, but it remains relatively large and clearly anamorphic}. It is useful to note that this design still remains a relevant example for newly developed optical systems  {such} as compact thermal imagers \cite{Alvarado22}. Therefore, it was taken as the basis for the demonstrator.

Once we have made the decision regarding the basic design geometry, we should choose the most appropriate basic design parameters. First, we explore the dependence of the image quality on the field and aperture. We presume that in any case the linear field of view (FoV), should correspond to the available commercial detector, i.e. $4096 \times 3072 \times 5.5 \mu m $ for the AMS CMV 12000 CMOS sensor \cite{CMV12000}, working with AXIOM camera  \cite{AXIOM}. We use the wavefront error (WFE) distribution across the field and aperture to compare different designs. To represent its variance across the aperture {,} we use the  {root-mean-square} (RMS) value over the pupil, which is calculated for an array of points uniformly covering the FoV. A typical distribution over the field  {represents} a well-corrected centre, which is easy to reach even with a limited number of variables, and a notable degradation towards the FoV corners, defined by field aberrations including the field curvature. Therefore, we use the maximum RMS WFE over  {all field points as a comparative metric} of the image quality. Following the spectral sensitivity of  {the} chosen sensor, we use $ \lambda=400 nm$ as the main reference wavelength  {because} it is close to the detector's working band minimum.  

Using this image quality criterion, we explore the $F/\# - FoV $ parameter space by changing their combination and re-optimizing the design. The region of interest in this  {parameter space} is defined as follows. The $F/\#$ varied  {in proximity to} that of the parent design. The upper aperture limit is set by the aberrations and beam clearances -- it appears that in designs faster than $F/\#=4.0$ the mirror tilts necessary to maintain the unobscured geometry become too large {,} and the aberrations grow rapidly. The lower limit is mainly defined  {by aberrations} -- in the case of an aperture slower than $F/\#=6.5$ the system becomes clearly diffraction-limited and further aberration correction makes little sense. In terms of \textit{FoV}, the region of interest is limited to the first extent by the detector and optics sizes. Starting with an estimation for the sensor format ranging from $2048\times2048$ with $10\mu m$ to $4096\times4096$ with $15\mu m$ pixels and  {the maximum diameter of optics equal to} $3 inches$ or $76.2 mm$, we obtain  {the} search range of $11-20^\circ$ for the angular field diagonal. However, as  {the aberrations} grow rapidly with the field extension, we crop this region to $11-17^\circ$.

\begin{figure}[htbp]
\centering
\fbox{\includegraphics[width=0.9\linewidth]{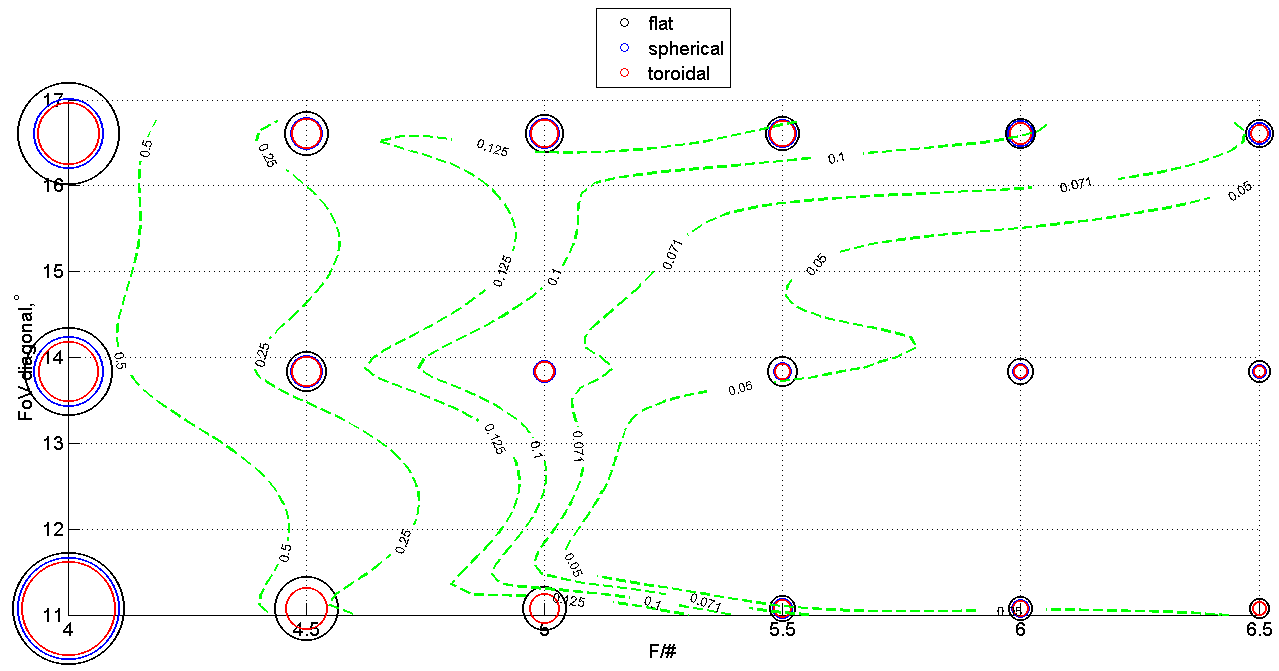}}
\caption{Maximum RMS WFE over the field in waves at 400 nm as a function of the $F/\#$ and diagonal of the angular field: circles show the values for individual designs, contour lines correspond to interpolation made to estimate the values in intermediate points.}
\label{fig:Shafer_WFEparam_space}
\end{figure}

In the optimization loop we use the mirror  { radii of curvature $R_i$, their tilt angles $\alpha_i$ and separation $d_i$ as the free variables}. In terms of boundary conditions, we limit the marginal ray incidence coordinates to keep  {the} clearance between the mirrors and the working beams. Also, we limit the mirrors clear aperture diameters to $74 mm$, the chief ray path length between two adjacent mirrors to $260 mm$ and the chief  {ray} angle of incidence at the detector centre to $7^\circ$.  {The weighted sum of geometrical aberrations and penalty terms representing the listed boundary conditions compose the merit function $f_{mer}$.}  We perform this exercise separately for the design with flat, spherical and toroidal detectors.  Fig.~\ref{fig:Shafer_WFEparam_space}  presents the output of this analysis. The circles correspond to separate optical systems developed in this loop and their diameter is proportional to the maximum RMS WFE. The dashed contour line illustrates an interpolated 2D distribution for the toroid-based designs obtained with these datapoints as input.  

\begin{figure}[htbp]
\centering
\fbox{\includegraphics[width=0.9\linewidth]{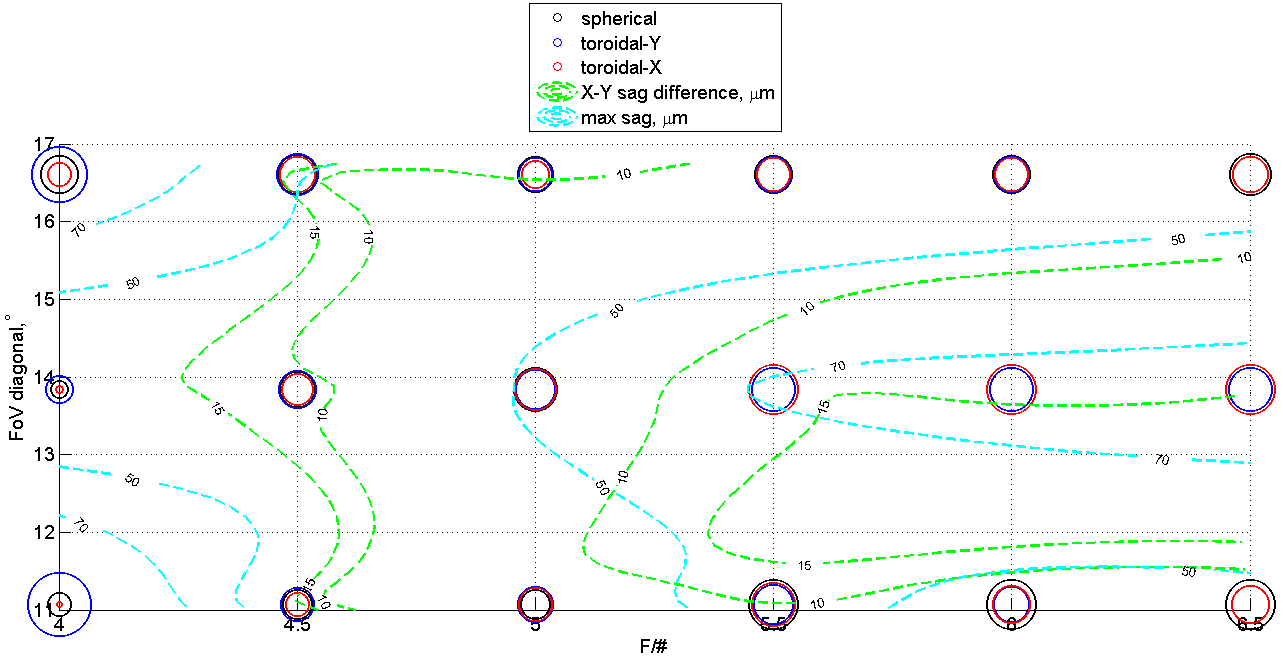}}
\caption{Maximum detector surface sag in microns as a function of the $F/\#$ and diagonal of the angular field: circles show the values for individual designs, contour lines correspond to interpolation made to estimate the values in intermediate points. }
\label{fig:Shafer_SAGparam_space}
\end{figure}

Another question, we would like to answer, before proceeding to building the demonstrator, is at which parameters combination the optimum detector shape anamorphism becomes significant. Therefore, we extract the maximum detector surface sag for each of the designs generated on the previous step and  {analyze} them in a similar way. Fig.~\ref{fig:Shafer_SAGparam_space} shows the sag for the spherical detectors and  {for the toroids} separately in X (sagittal) and Y (tangential) directions. Similarly, interpolated contour lines are added to guide the eye - the maximum sag of toroidal detectors and the maximum difference in the two sections are plotted.

There are few points to note in the  {two diagrams exploring the parameter space}. First, as we could expect, the required detector curvature and the residual WFE, grow towards faster designs, but at $F/4$ both values raise sharply.  { For the focal ratios slower than $F/5$ the absolute image quality remains relatively good with $RMS~WFE<\lambda/10$. However, the image quality difference conditioned by the detector shape remains measurable in this zone of the parameter space}. In terms of the surface sag both the absolute sag and anamorphism have the minimum. It would be desirable to work with values of tens of  {micrometers} to be confidently an order of magnitude above the manufacturing and measurement errors. So, the right zone outlined by the $10 \mu m$  {dashed line} is of the highest interest.

In terms of the variation with the field, the absolute WFE reachable with the toroidal detector changes slowly with the  { maximum field angle, as one can see from the circle diagrams in columns, or long vertical sections of the contour lines. At the same time {,} at slower focal ratios, the WFE distribution exhibits a minimum around $13^\circ$.} The difference from flat and spherical designs becomes more apparent either in this minimum zone or at higher apertures and fields with the growth of absolute values. From the surface shape standpoint the absolute sag and its anamorphic difference grow at medium field values of $13-14^\circ$ with low $F/\#$'s, or at extreme fields with the highest $F/\#$'s.

A few comments should be made to explain the observed non-linearities in the distributions of these metrics. First, as we fix the linear field size and change the angular field size, the effective focal length changes between the systems {,} and one should keep this in mind. Second, the multiple geometrical boundary conditions introduced to avoid cropping the working beam and collisions between the components make a major effect on the local minima search. Finally, under some circumstances other aberrations  {as} astigmatism can dominate {,} and then varying the image surface curvature becomes inefficient {,} thus causing its fast changes in the numerical optimization loop.

\begin{figure}[htbp]
\centering
\fbox{\includegraphics[width=0.55\linewidth]{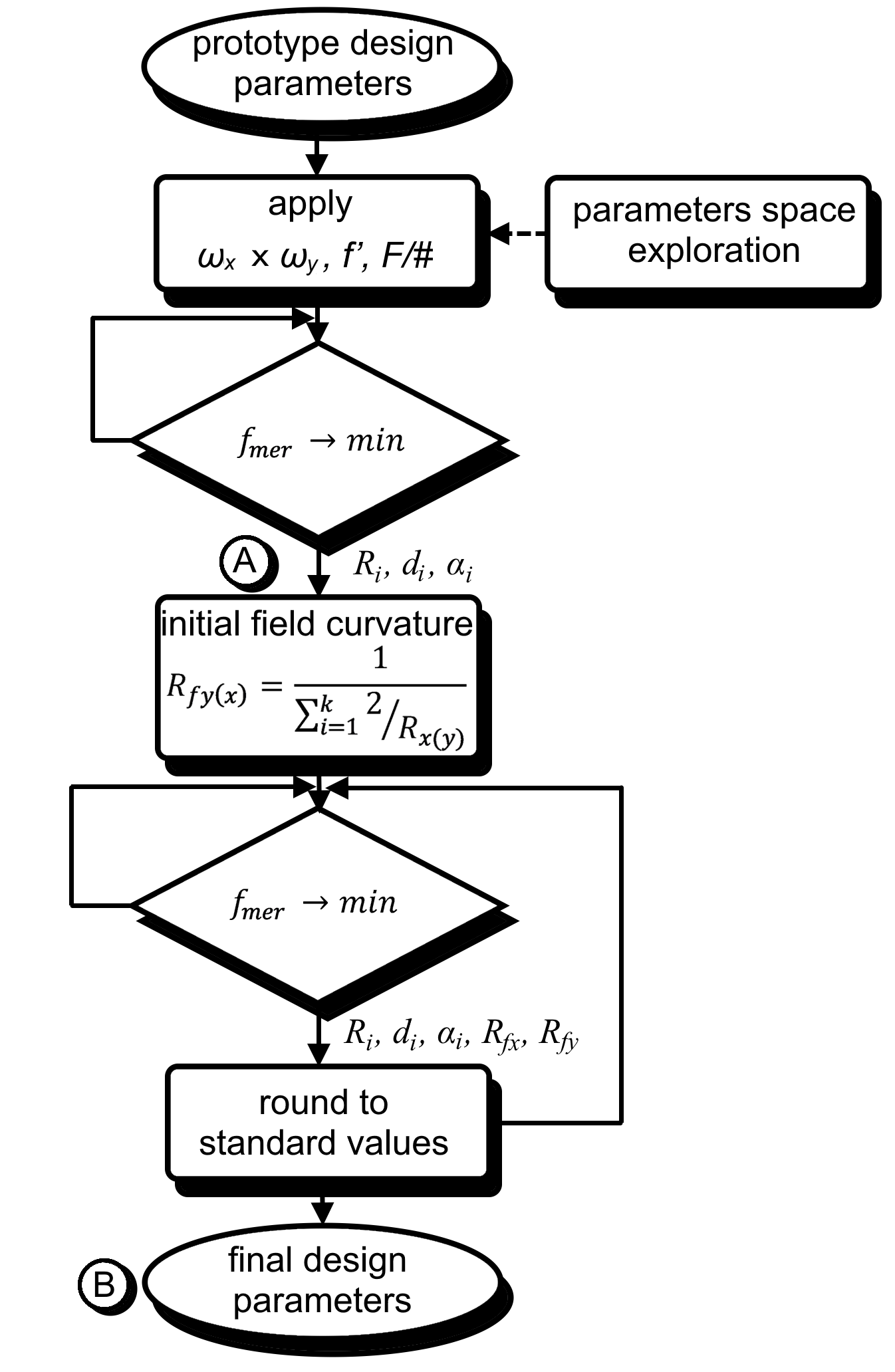}}
\caption{Simplified flowchart of the telescope design algorithm.}
\label{fig:Shafer_algorithm}
\end{figure}

Taking into account all of the listed considerations, we decided to use the following configuration for the demonstrator's optics: the effective focal length is  {$f'=121.2~mm$, $F/\#$ is 5.5, the angular FoV is $\omega_x \times \omega_y = 10.6^\circ \times 8^\circ$}, corresponding to the linear FoV is $22.5~mm \times 16.9~mm$. 
We re-optimized the design with fixed radii of the concave mirrors (\textit{M2} and \textit{M4}) to be able to use off-the-shelf concave mirrors. Also, additional boundary conditions were introduced to increase the distances between components and allow  {use of} standard optical mounts. The anamorphic curvatures of the focal surface   {$R_{fx}$ and $R_{fy}$  }at the starting point were defined by the method described before  \cite{Muslimov18}. Then we specified these values in the numerical optimization loop (see Fig.~\ref{fig:Shafer_SAGparam_space}). After  {fine-tuning the design} to fit the commercial components {,} we obtained the following curvatures: $614mm$ in the tangential plane and $919mm$ in the sagittal one. The full optical prescription of the telescope is given in Table~\ref{tab:prescr}, so any interested reader can reproduce the design and extend the analysis or use it as a starting point for a new application.   Note that the stop diaphragm is set on \textit{M2} surface.  {The data in this table appears in the format used in optical design software, so the sign inversion after reflection is included.}
{The design and optimization sequence is also shown as a flowchart in Fig.~\ref{fig:Shafer_algorithm} to facilitate repeating our design.}

\begin{table}
    \centering
     \caption{Optical prescription of the telescope for prototyping}
    \begin{tabular}{|p{1.2cm}|p{2.5cm}|p{2.5cm}|p{2.5cm}|p{2.5cm}|}
    \hline
    Surface	&Radius of curvature, mm &	Centre-to-centre distance, mm	& X tilt angle, $^\circ$  &	Clear        aperture, mm \\
\hline
M1  {-CX}	&249.227&	-239.212&	12.445&	$	\varnothing 41.6$\\
\hline
M2  {-CC}	&406.400&	201.032&	-7.6&	$	\varnothing 67.3$\\
\hline
M3	 {-CX} &704.200&	-183.783&	14.813&	$	\varnothing 52.8$\\
\hline
M4 {-CC}	&406.400&	161.998&	-7.717&	$	\varnothing 69.0$\\
\hline
det.  {-CC}&	-614.06/-919.42&	-&	-2.511&	22.5x16.9\\
\hline
    \end{tabular}
   
    \label{tab:prescr}
\end{table}

\begin{figure}[htbp]
\centering
\fbox{\includegraphics[width=0.75\linewidth]{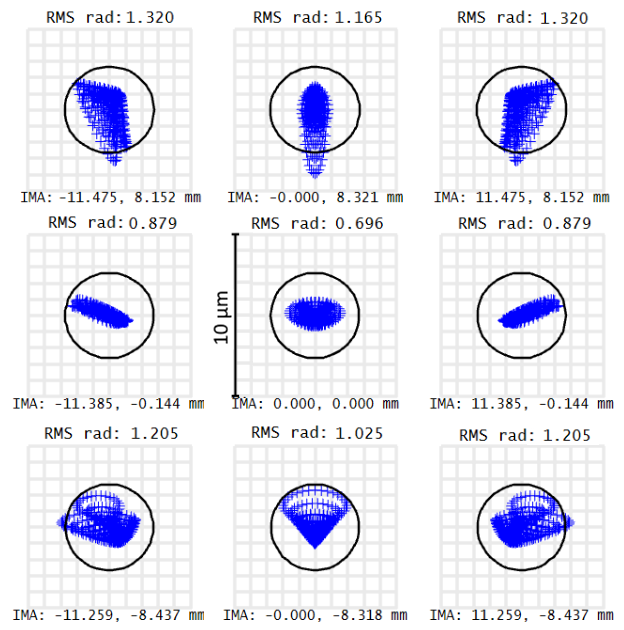}}
\caption{Spot diagrams of the unobscured telescope with toroidal detector compared to the Airy disk at 400nm.}
\label{fig:Shafer_Spots}
\end{figure}

In order to provide a general impression of the aberrations correction in the design {,} we plot its spot diagrams in Fig.~\ref{fig:Shafer_Spots}. As the diagrams show, the design is practically  {diffraction-limited} at the shortest wavelength over the entire field.  

\begin{figure}[htbp]
\centering
\fbox{\includegraphics[width=0.55\linewidth]{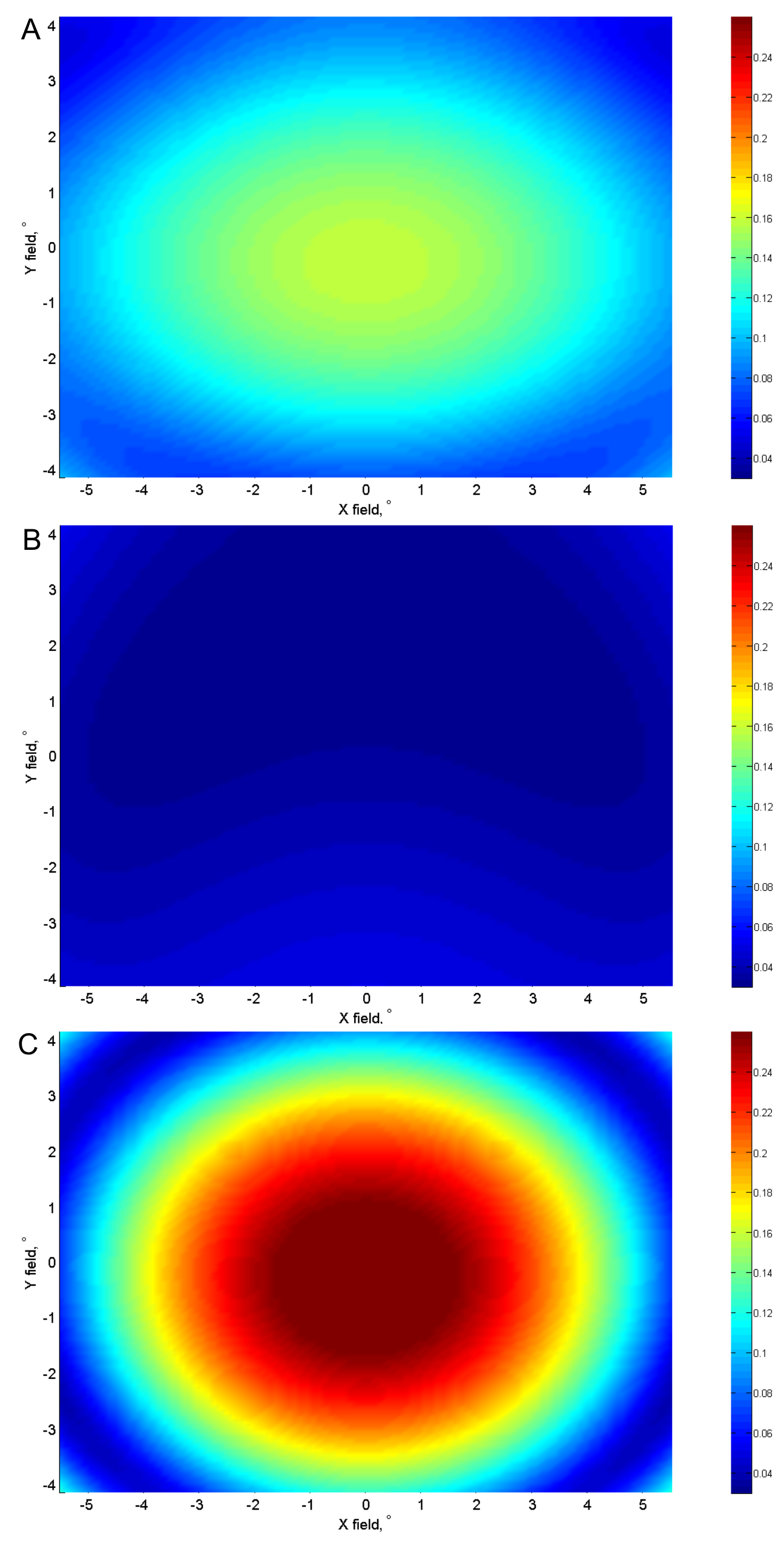}}
\caption{RMS WFE distribution over the field: A -- telescope designed and operated with a flat detector, B -- telescope designed and operated with a toroidal detector, C -- telescope designed with a toroidal and operated with a flat detector. }
\label{fig:Shafer_WFEmaps}
\end{figure}

 {There are a few other questions that} we should answer with modeling before we can move on  {to manufacturing} and measurements. First, it would be desirable to perform a comparative study for different detector shapes. However, as we saw before, the optimization converges to different solutions depending on the presumption regarding the detector shape taken  {at} the beginning. At the same time, it would be quite difficult in practice to build and characterize two separate optical systems with different sensors. The question is: can we use the same optical system with flat and curved detectors and what would be the expected difference in the image quality? To illustrate the difference {,} we compare three cases (see the corresponding wavefront maps in Fig.~\ref{fig:Shafer_WFEmaps}): \textit{A} -- the design was optimized for a flat sensor and is operated with the flat sensor as well, \textit{B} -- the design was optimized for a toroidal sensor and is operated with such; \textit{C} -- a flat sensor is substituted into the design \textit{B} and its longitudinal position and tilt are re-optimized.  {Note that the \textit{A} and \textit{B}  {designs} are shown also as outputs in Fig.~\ref{fig:Shafer_algorithm}}.  

Table~\ref{tab:WFE_maps} summarizes the numerical estimates for the WFE distribution over the FoV in each of the cases described above. As one can see, the design \textit{A} initially optimized for a flat detector exhibits a better performance than design \textit{C} with a flat sensor substituted at a later stage. However, the toroid-based design \textit{B} is superior over both of them {,} and the difference is much more notable -- the gain in median WFE is $2.1$ times higher. Therefore, we conclude that the design \textit{C} is to a certain extent representative for the image quality studies and will use it in a comparative experiment.

\begin{table}
    \centering
    \caption{Numerical estimates of the WFE distribution across the field for the experimental designs.}
    \begin{tabular}{|l|c|c|}
    \hline
         Design & Max. WFE, waves at 400 nm  & Median WFE, waves at 400 nm\\
         \hline
         A &  0.1591 & 0.1074 \\
         \hline
         B &  0.0537 & 0.0351 \\
         \hline
         C &  0.2766& 0.1534 \\
         \hline
    \end{tabular}
    
    \label{tab:WFE_maps}
\end{table}
 
Finally, we should estimate the required tolerances to demonstrate that the expected image quality is reachable with practically feasible error margins and that adding the new design variables, as the two curvatures of the image surface, has a minor impact on the overall errors budget. We use the specifications of commercial mirrors  \cite{Edmund} and  {sources from the literature}   \cite{Yama23} as a reference for individual tolerances estimation. We  {set} the weighted average value of as-built RMS WFE to be below $1\lambda$ at 400 nm as the tolerance analysis criterion. Using the numerical derivatives, we obtain the initial tolerances and truncate them to the nearest technologically feasible values. Then we tune the entire set of tolerances to achieve $90\%$ confidence over 100 Monte-Carlo runs. The results for the baseline design with a toroidal detector  {(i.e. design \textit{B})} are shown in Table~\ref{tab:tolerance}. For comparison {,} we perform the same analysis for the version based on a flat sensor  {(design \textit{A})}. The results are very similar. Only the  {values, which have} some notable deviations {,} are given in  {parentheses} and highlighted in red in Table~\ref{tab:tolerance}. In general,  {all the tolerances} are reachable. The only exclusion is the M2 surface shape - as it is located at the system's pupil, it appears to be more sensitive to the sag deviations. The resultant tolerances for M2 are tighter than the precision guaranteed for an off-the-shelf component. This could be solved either by measuring the real component and correcting the design prior to  {its} assembly, or just by mitigating the target image quality. Another point to note here is that the tolerances almost do not change when re-optimizing the design for a toroidal detector. In the cases when the difference is notable, the tolerances  {appear to be tighter rather than looser for the flat detector} as the nominal  design  {\textit{B}  with toroidal detector} is corrected better.

\begin{table}
\caption{Tolerances estimates for the telescope: black -- values calculated for the toroidal detector design \textit{B}, red -- values discrepant for the flat detector design \textit{A}.}
    \centering
    \begin{tabular}{|p{2.95cm}|p{1.6cm}|p{1.6cm}|p{1.6cm}|p{1.6cm}|p{1.6cm}|}
    \hline
         Component & M1 & M2 & M3 & M4 & Detector \\
         \hline
         Radius, $\%$ & 2.0 & 0.6 & 1.9 \textcolor{red}{(1.7)} & 2.0  & 2.0/2.0\\
         \hline
         RMS irregularity, waves at 546 nm & $\lambda/4$ \textcolor{red}{($\lambda/5$)} & $\lambda/7$ \textcolor{red}{($\lambda/8$)} &  $\lambda/4$ \textcolor{red}{($\lambda/5$)}  & $\lambda/5$ & $4\lambda$ \\
         \hline
         Thickness, mm & 0.5 & 0.5 & 0.5 & 0.5 & - \\
         \hline
         Tilt, arcmin & 6 & 2 & 6 & 6 \textcolor{red}{(8)}  & 96\\
         \hline
    \end{tabular}
    
    \label{tab:tolerance}
\end{table}

\section{Toroidal detectors}
\label{sec:detector}
For the lab prototype  {of telescope} we used an AMS CMV12000 sensor as described above. It was proven to be suitable for science-grade instruments in terms of dark current and readout noise and to maintain their properties after bending \cite{Lombardo19}.

\begin{figure}[htbp]
\centering
\fbox{\includegraphics[width=0.8\linewidth]{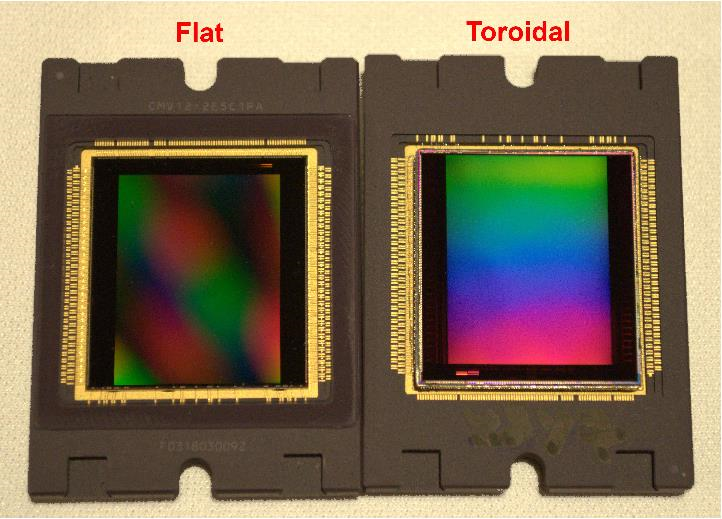}}
\caption{Two CMOS sensors used in the experiment : left -- flat, right -- toroidally curved.}
\label{fig:Shafer_sensors}
\end{figure}

This detector was curved down to the calculated toroidal shape by means of  {the technology} used in earlier studies  \cite{Jahn17, Gaschet19, Chambion18, Lombardo19}. The technology was slightly modified by changing the mechanical template used in the bending facility to generate different curvatures in two directions. The toroidal detector manufactured with it is shown in Fig.~\ref{fig:Shafer_sensors} on the right, next to the initial flat one. One may note that the reflected images of the spot lamps are turned into stripes, since the curved surface acts as an anamorphic mirror.  

\begin{figure}[htbp]
\centering
\fbox{\includegraphics[width=0.8\linewidth]{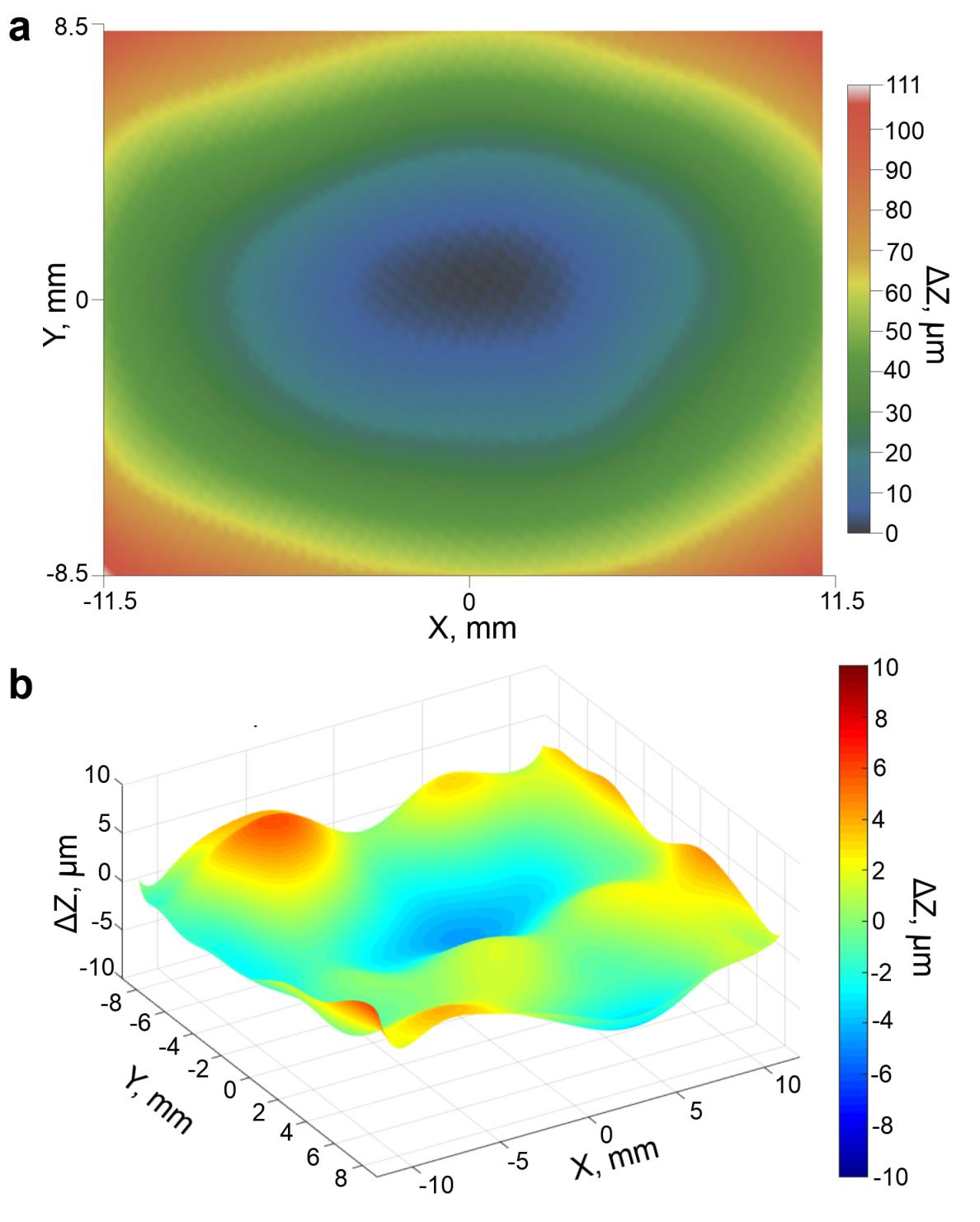}}
\caption{The toroidal detector surface shape: a -- the measured shape, b -- the deviation from calculated shape (RMS = 2.4$\mu m$, PTV = 10.8 $\mu m$).}
\label{fig:Shafer_toroid}
\end{figure}

The generated detector surface shape was verified by profilometry. The exact measured shape is presented in Fig.~\ref{fig:Shafer_toroid}~a. One may note  {the difference of curvature} in two directions. The residual deviation of the produced detector surface shape from the calculated one is shown in Fig.~\ref{fig:Shafer_toroid}~b. The deviation from the calculated shape is  $10.8 \mu m$ peak-to-valley (PTV) or $2.4 \mu m$ RMS. These values are significantly smaller than the depth of focus, which is at least $\pm 30 \mu m$ for the given optical system parameters and below the approximate tolerance value shown in Table~\ref{tab:tolerance}. So, the plot demonstrates that the surface shape is reproduced with a sufficiently high precision.

\section{Lab prototype and experiments}
\label{sec:experiments}

\begin{figure}[htbp]
\centering
\fbox{\includegraphics[width=0.9\linewidth]{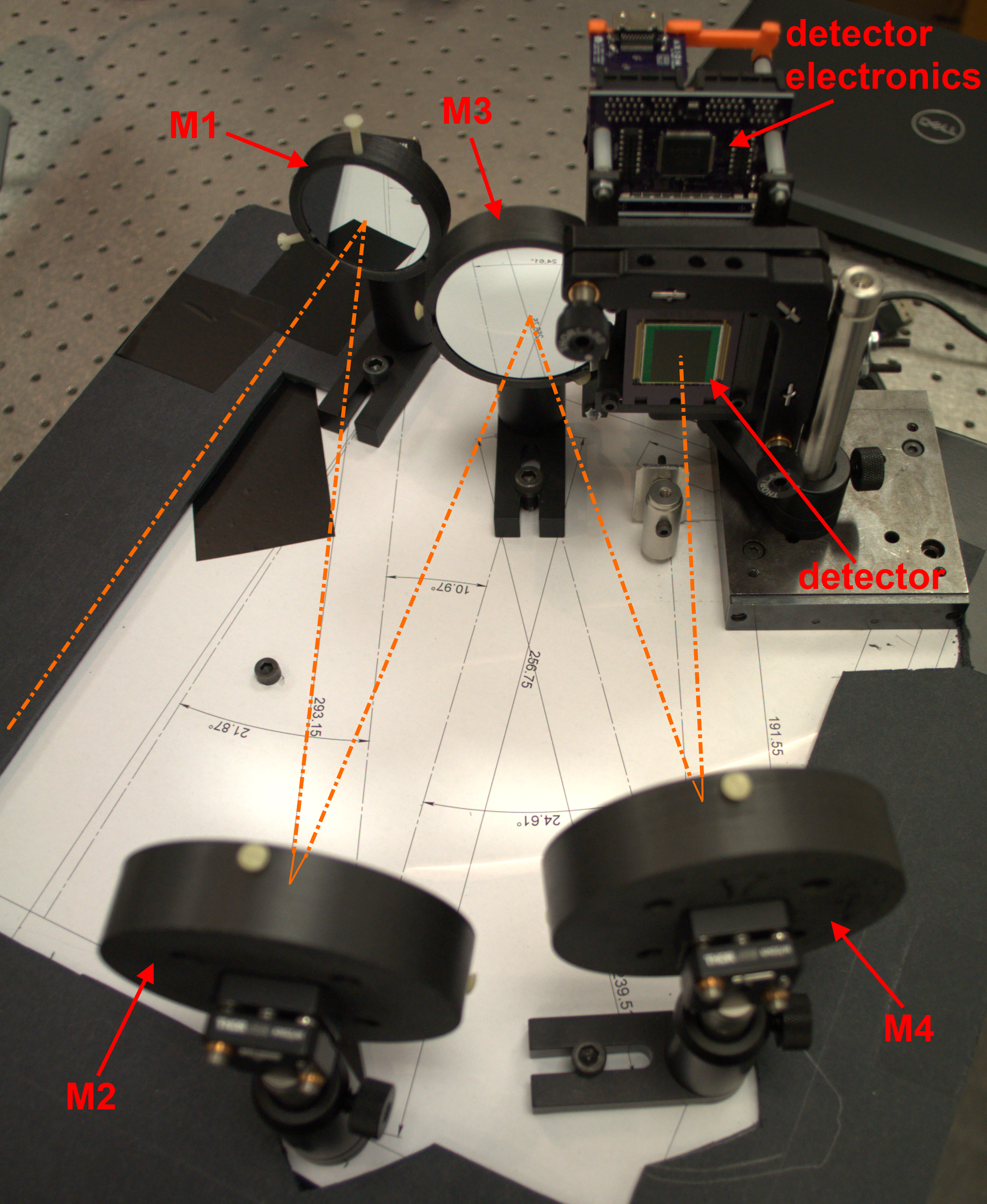}}
\caption{ {Lab prototype of the off-axis telescope with toroidal detector.}}
\label{fig:Shafer_setup}
\end{figure}

On the basis of the developed optical design {,} a telescope lab prototype was built. The four mirrors are installed in 3D-printed plastic holders attached to standard adjustable mounts and the entire optical system is assembled on a breadboard. The assembled telescope prototype is presented in Fig.~\ref{fig:Shafer_setup}.

The detector was operated with an open-source Apertus-AXIOM camera  \cite{AXIOM} {.} The entire unit was installed in a mount with 3 degrees of freedom -- tip, tilt {,} and focus.

\begin{figure}[htbp]
\centering
\fbox{\includegraphics[width=0.9\linewidth]{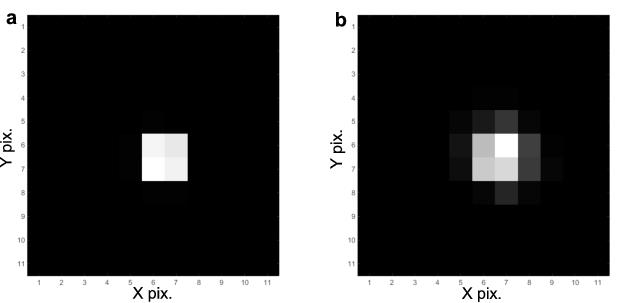}}
\caption{Point source image in the FoV center: a --- modelling, b --- experimental measurement.}
\label{fig:Shafer_PSF}
\end{figure}

For the imaging performance tests we used a telephoto lens ($f'=300~mm, F/\#=5.6$) as  {a} collimator. The tests were performed in two modes with two different light sources. 
 For the alignment and the first test we used a point source, represented by a single mode fiber connected to a Fabry-Perot benchtop laser source operating at $635~nm$. 
We used this source to estimate the point spread function (PSF) of the telescope in FoV centre. Here we must note that we used software pixel binning to implement the monochrome mode in the RGB sensor, so the actual resolution element size was $11 \times 11 \mu m^2$. For comparison {,} we  {modeled} the same image accounting  {for aberrations and sampling}. The results are shown in Fig.~\ref{fig:Shafer_PSF}. The size and shape of the image obtained by  {modeling} and measurements are  {in good} agreement.

\begin{figure}[htbp]
\centering
\fbox{\includegraphics[width=0.6\linewidth]{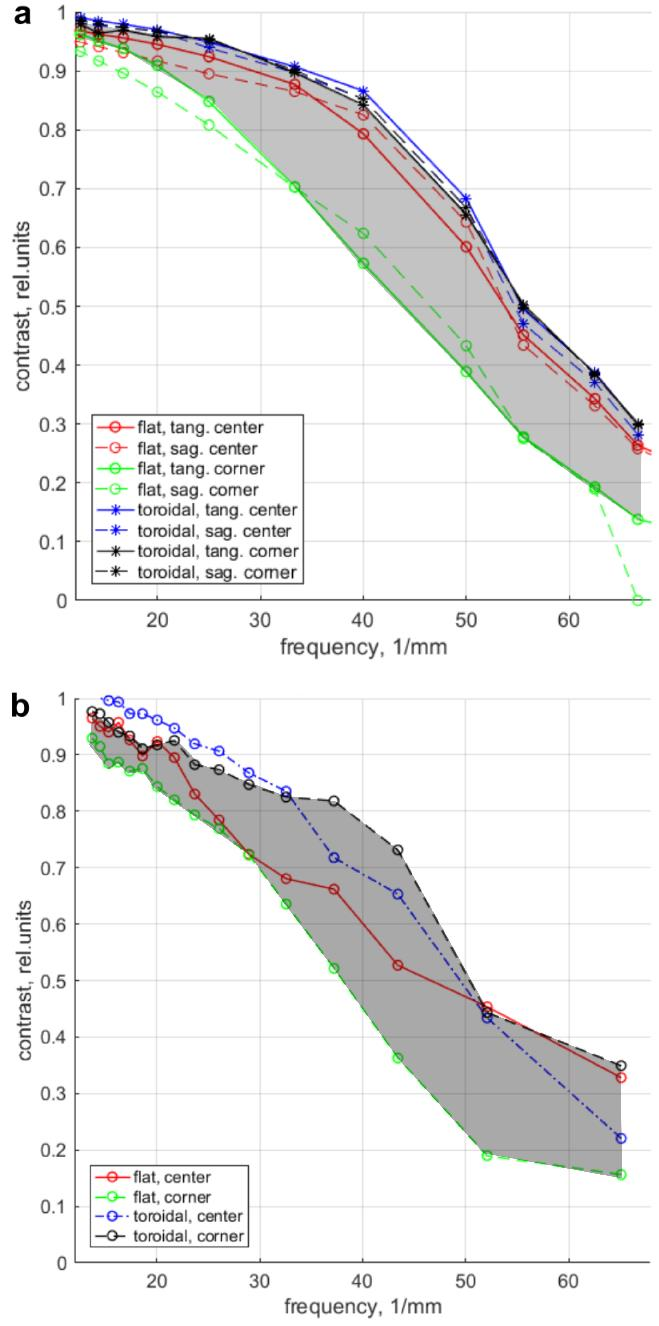}}
\caption{Telescope effective MTF: a --- simulation, b --- experimental measurement.}
\label{fig:Shafer_MTF}
\end{figure}

In the second test mode {,} we used an extended source represented by organic light emitting diode (OLED) microdisplay  {with} active area of $1280 \times 1024$ pixels and $12 \mu m$ pixel size. A standard radial test chart was projected  {onto} the display and imaged by the telescope prototype. Then the image was divided into annular zones and for each zone the image contrast was computed. The resultant dependence of the contrast on spatial frequency represents the telescope's effective module transfer function (MTF). The measurements were performed for two points in the FoV -- the center  {(magenta beam in Fig.~\ref{fig:Shafer_design})} and the   {top left corner (orange beam in Fig.~\ref{fig:Shafer_design}). As one can note from Fig.~\ref{fig:Shafer_Spots}, the aberrations on the left and right sides are symmetrical due to the design symmetry with respect to the tangential plane, while those on the top and bottom sides are very close to each other. So, we conclude that one off-axis point is enough to characterize the imaging performance, while adding more reference points would only complicate reading of the MTF plots.} 
In order to estimate the imaging performance gain achieved due to use of the anamorphic curved detector, we repeated the same measurements with the initial flat sensor (see design \textit{C} in Sec.~\ref{sec:design}). 
Similarly, for comparison the effective MTF with the same sampling was obtained by means of  {modeling}. All  {of} the MTF curves are given in Fig.~\ref{fig:Shafer_MTF}.

The plots show good agreement between the computations and the experiment, both qualitative and quantitative. And, more important, both diagrams demonstrate the gain in contrast obtained with the toroidal detector over the extended FoV. For instance, the predicted contrast compared to that of the telescope with  {a} flat detector at $50 ~mm^{-1}$ is higher by $0.27$, while the experimentally measured difference is $0.29$, i.e. the difference is $7.4\%$. In general, the deviation between the measured and simulated contrast gain is $0.04$ units RMS.
 Thus {, testing} of the lab prototype fully confirms the functionality of the toroidal curved detector and its advantages in terms of image quality improvement.  

 \section{Conclusion}
\label{sec:conc}

In this work {, we have developed} a four-mirror unobscured telescope design with a toroidal detector {,} studied its properties {,} and built a lab prototype on its basis. This work has the following outcomes, which we think could be of  {a certain} interest to the optical instrumentation community. 

\begin{enumerate}
    \item To the best of our knowledge, this is the first experimental demonstration of the application of an anamorphically curved detector surface. The PSF size and MTF shape measurements are in good qualitative and quantitative agreement with the modeling predictions. This means that with the current bending technology the concept of anamorphically curved sensor can be implemented in practice with  { a} sufficient precision - down to $2.4 \mu m $ RMS surface error. Expanding our results further, we can expect that the detector surface shape is limited by its steepness and the corresponding mechanical stress, rather than by the number of variables used to describe the surface.
    \item The design parameter space study shows that use of a toroidal detector in a 4-mirror unobscured design with $F/\#=4..6.5$ and $FoV=11..17^\circ$ leads to a gain of  up to $1.14\lambda$ over a flat detector and up to  $0.26 \lambda$ over a spherical detector in terms of \textit{RMS WFE}. This provides a convenient merit  {for} the expected gain in imaging performance and can be used for other designs of a similar type in order to estimate {,} if introduction of an anamorphically curved detector is rational. 
    \item The developed 4-mirror telescope optical design can be of a certain interest by itself. It  {shows} an excellent performance, providing diffraction-limited image over the large FoV with $13.3^\circ$ diagonal by design. The telescope consists only of four standard spherical concave mirrors and is free of obscuration. Even using standard lab mounts the overall size of the optical system is $\approx 321 \times 268 \times 135 mm^3$ and it could be reduced further with use of custom-machined mechanics. 
\end{enumerate}

The success of this experimental study opens new prospects  {for} design of optical systems with curved focal surface arrays.  We can propose a number of applications, for which our approach can be useful:
\begin{itemize}
    \item Building of unobscured off-axis optical systems and upgrade of the existing ones similar to past space missions \cite{Hammar19, Martin19}. Special attention should be paid to space applications, where such features of the proposed design as high compactness (i.e. high ratio between the collecting area and overall volume), high image quality over the extended field, wide working spectral range limited only by the  {reflective} coating and the detector sensitivity can become the decisive factors. Some fly-by space missions payloads can serve as good examples  \cite{Liu24}.
    \item Use in optical systems working with anamorphic beams, for example image slicers used in integral-field spectrographs  \cite{Henault03, Thatte22}.  
    \item Use in optical systems imaging a curved object of  {complex shape}, both natural  \cite{Hillman07, Racicot22}  and artificial  \cite{Chektybayev2015} ones.
     
\end{itemize}

 {Obviously, the efficiency of anamorphic detectors application will also depend on the optical design type. For instance, as the field curvature grows with the field angle, it may be desirable to use such curved detectors in systems operating with  {a} wide field of view of a high aspect ratio \cite{Schifano22}. However, this approach presumes that the rest of the aberrations across the field are well corrected, which may be reached with freeform optics. In addition, the curvature along the short side of a rectangular field of view may be negligible, which would lead to a cylindrical focal surface representing a particular and simplified case of our solution. For the two-mirrors unobscured designs as Yolo or off-axis Gregory telescopes we do not expect our approach to be particularly efficient as the field curvature is relatively small and does not make any significant contribution to the image quality. This applies both to the classical  \cite{Buchroeder72, Yin21} and advanced  \cite{Bazhanov17, Zhong17} versions of these designs. At the same time we believe that the anamorphic focal surface approach may be useful for a large variety of three mirror anastigmat (TMA) telescopes \cite{Liu24,Greggio16}. Re-design of these telescopes for operation with anamorphically curved detectors can become a subject of separate study. }

\subsection* {Acknowledgments}
The authors would like to acknowledge the European commission for funding this work through the Program H2020-ERC-2015-STG–678777 - ICARUS of the European Research Council (all Authors) and through the program H2020-ERC-2018-POC-839271 - CURVE-X (T.B. and E.H.).

J.L. thanks for the support of China Scholarship Council (CSC).

The authors also would like to warmly thank their former colleague Dr. Kelly Joaquina, who contributed to the curved-sensor camera commissioning in 2019-2021.  

 {The authors also would like to thank their colleague, Dr. Kjetil Dohlen from LAM for his help with the final manuscript preparation}.


\bibliography{article}   
\bibliographystyle{ieeetr}   

\end{document}